\documentclass[10pt]{iopart}
\usepackage{epsfig,cite}
\def\lsim{\raise0.3ex\hbox{$<$\kern-0.75em\raise-1.1ex\hbox{$\sim$}}}
\def\gsim{\raise0.3ex\hbox{$>$\kern-0.75em\raise-1.1ex\hbox{$\sim$}}}

\newcommand{\bec}{\begin{center}}
\newcommand{\enc}{\end{center}}
\newcommand{\beq}{\begin{equation}}
\newcommand{\eeq}{\end{equation}}

\newcommand{\beqar}{\begin{eqnarray}}
\newcommand{\eeqar}{\end{eqnarray}}

\begin{document}

\title{
      Influence of jets and resonance decays on the constituent quark
      scaling of elliptic flow }

\author{
E~E~Zabrodin\dag\ddag, L~V~Bravina\dag, G~Kh~Eyyubova\dag\ddag, 
I~P~Lokhtin\ddag, L~V~Malinina\dag\ddag\S, S~V~Petrushanko\ddag, 
A~M~Snigirev\ddag  
}  
\address{\dag\
         Department of Physics, University of Oslo, Oslo, Norway}
\address{\ddag\
         Skobeltsyn Institute of Nuclear Physics, Moscow State 
         University, Moscow, Russia}
\address{\S\
         Joint Institute for Nuclear Researches, Dubna, Moscow
         Region, Russia}

\begin{abstract}
The Monte Carlo HYDJET++ model, that contains both hydrodynamic
state and jets, is applied to study the influence of the interplay
between soft and hard processes on the formation of the elliptic 
flow in heavy-ion collisions at RHIC and LHC energies. Jets are
found to cease the hydro-like increase of the elliptic flow with
rising $p_T$. Since jets are more influential at LHC than at RHIC, 
the $v_2$ at LHC should be weaker than that at RHIC. Violation of 
the number-of-constituent-quark (NCQ) scaling is predicted. The
decays of resonances are found to enhance the low-$p_T$ part of
the elliptic flow of pions and light baryons, and work toward the
fulfillment of the NCQ scaling.
\end{abstract}




\section{Introduction}
\label{sec1}

The elliptic flow of hadrons, produced in the course of non-central
heavy-ion collisions, is considered as one of the main probes of the
reaction early stage. It originates from the rescattering process
in the initially anisotropic overlapping region. Recall that elliptic
flow is the second coefficient of Fourier decomposition of particle
invariant cross section in the azimuthal plane, namely
$v_2 = \langle \cos{(2\phi)} \rangle \equiv \langle (p_x^2 - p_y^2)/
p_T^2 \rangle$. Here $\phi,\ p_T,\ p_x,\ {\rm and}\ p_y$ are the 
azimuthal angle, particle transverse momentum, and its in-plane and 
out-of-plane components, respectively. The spatial anisotropy of the 
almond-shaped area is then transformed to the anisotropy in the 
momentum space. When the isotropy in the coordinate space is restored, 
the elliptic flow stops to develop anymore. Thus $v_2$ is linked to 
the equation of state (EOS) of hot and dense partonic matter formed in 
the very beginning of the collision.

Many features of the flow at energies up to top RHIC energy $\sqrt{s}
= 200$\,AGeV are successively reproduced by several models, in
particular, hydrodynamic models (for review, see \cite{VPS08} and
references therein). However, predictions of these models for the
elliptic flow at top LHC energy $\sqrt{s} = 5.5$\,ATeV often 
contradict each other. For instance, simple scaling model \cite{Last}
favors increase of $v_2$ at midrapidity to $v_2(y=0) \approx 7-8\%$,
several models predict saturation of the flow \cite{Last}, whereas 
ideal hydro indicates \cite{KH09,NER09} that $p_T$-integrated flow at 
LHC will be larger than that at RHIC although $v_2^{\rm LHC}(p_T) 
\leq v_2^{\rm RHIC}(p_T)$ in the domain of low and intermediate 
transverse momenta.

Jets contribute to the elliptic flow via the so-called jet quenching
mechanism that accounts for the quite weak flow of about 5\%.
The simultaneous description of the flow caused by the soft processes
and the jets, and study of the interplay between these two mechanisms
has been done recently in \cite{hyd_v2_prc09} by means of the HYDJET++
model \cite{hydjet++}. The model contains the soft physics part, 
represented by the parametrized hydrodynamics with given freeze-out
conditions \cite{fastmc}, and the jets. The production of hard 
multiparton states and their propagation in dense partonic matter
takes into account radiative and collisional losses \cite{hydjet}
associated with the parton rescattering in the expanding QGP. The
contribution of both soft and hard parts to the total multiplicity of
a heavy-ion collision at ultrarelativistic energies depends merely on 
the collision energy and centrality. The few free parameters of the 
model have been tuned to provide adequate description of hadron 
spectra, elliptic flow, femtoscopic momentum correlations, and high 
$p_T$ tails of hadron spectra in gold-gold collisions at RHIC 
energies. Further description of the HYDJET++ model can be found in
\cite{hydjet++,fastmc,hydjet}. The main aim of our present paper is 
the study of the interplay between soft and hard processes responsible
for the elliptic flow production in heavy-ion collisions at energies
of RHIC and LHC, respectively. The influence of the resonance decays 
on the final elliptic flow of particles is very important at both 
energies in question. Here, we are going to benefit from the rich 
table of baryon and meson resonances (about 360 states) implemented
in the HYDJET++ model.   
   
\section{The influence of jets and resonances on the $v_2$ formation}
\label{sec2}
 
\begin{figure}[htb]
\epsfig{file=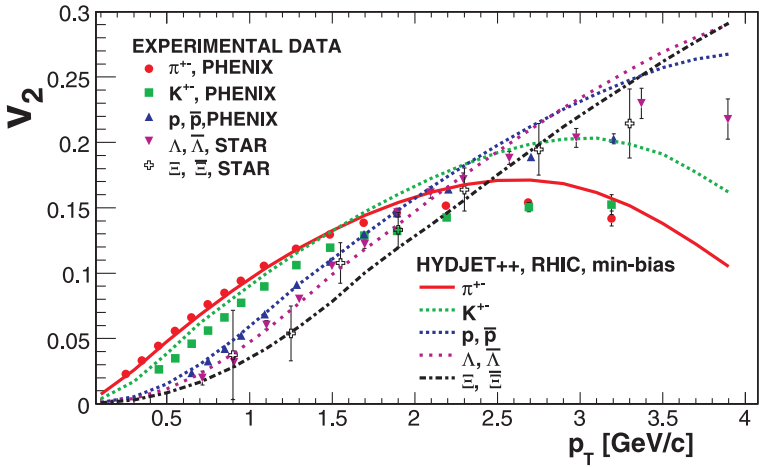,width=100mm}
\caption{
The $v_2(p_T)$ dependence in the HYDJET++ model for different 
hadron species (lines) and comparison with RHIC data (symbols).
}
\label{fig1}
\end{figure}

The model description of the particle elliptic flow as a function
of the transverse momentum is plotted in Fig.~\ref{fig1} onto the
RHIC experimental data taken at $\sqrt{s}=200$\,AGeV. Besides the
standard rise of the flow excitation functions and the mass ordering
of the particle flow at $0 \leq p_T \leq 2$\,GeV/$c$, the model 
exhibits also the saturation and the drop of the particle $v_2(p_T)$ 
at $p_T \geq 2.5$\,GeV/$c$ accompanied by the violation of the mass 
ordering. The last two features cannot be attributed to ideal 
hydrodynamics that indicates the instant rise of the elliptic flow up 
to unity without the crossing of meson and baryon branches as $p_T$ 
rises to infinity. Here we see the result of the interplay between the
soft hydro processes and hard jets. In contrast to hard processes, the 
multiplicity of hadrons produced in soft processes drops exponentially 
with increasing transverse momentum. Therefore, after a certain $p_T$ 
the particle spectrum is dominated by the jet particles. The latter 
carry quite weak elliptic flow, thus the combined flow of high $p_T$ 
hadrons drops. Also, the slope of the $p_T$ spectra of heavy hadrons
is not as steep as that of the light particles. Because of this, the
hydro component of the transverse momentum distribution of heavy
particles dominates until larger values of $p_T$ resulting to the 
change of the mass ordering of the flow: After a certain $p_T$ the
heaviest particles possess the largest flow.   

\begin{figure}[htb]
\begin{minipage}[t]{67mm}
\epsfig{file=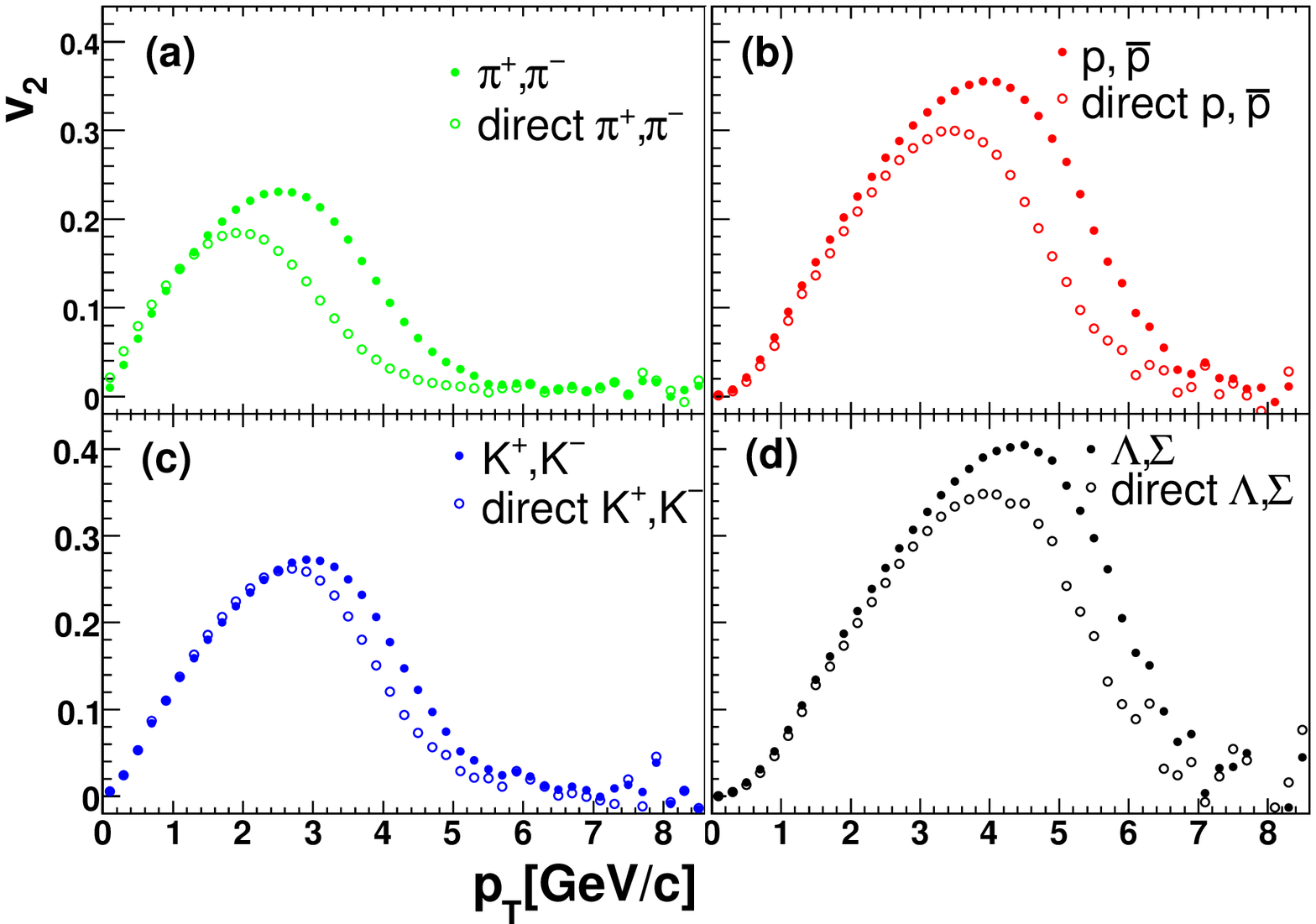,width=67mm}
\end{minipage}
\hspace{\fill}
\begin{minipage}[t]{67mm}
\epsfig{file=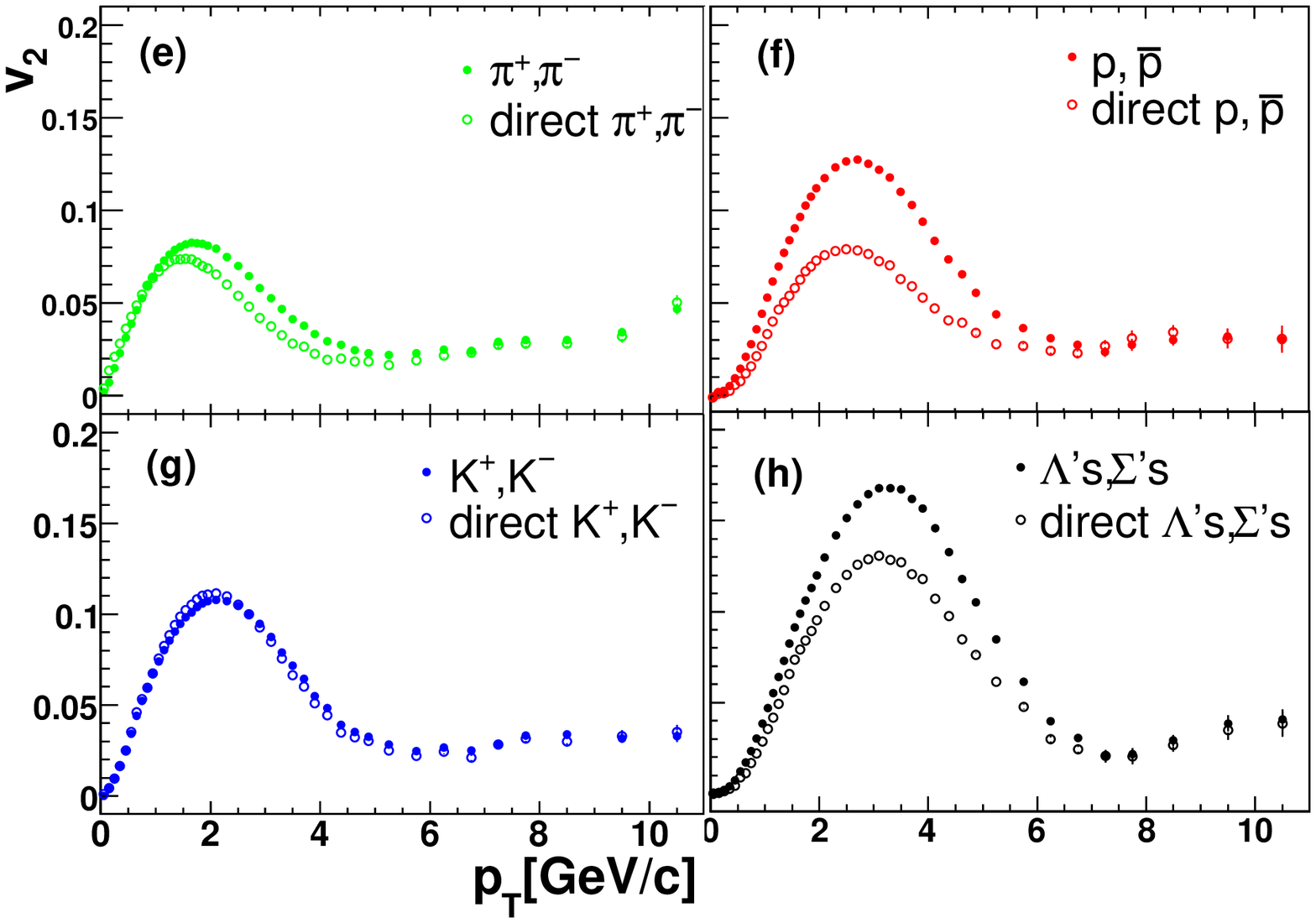,width=67mm}
\end{minipage}
\caption{ Left:
The $v_2(p_T)$ dependences of all hadrons (full symbols) and of
direct hadrons (open symbols) in the HYDJET++ simulations of
Au + Au at $\sqrt{s}=200$\,AGeV with centrality $c = 42\%$: 
(a) $\pi^{ch}$, (b) $p+\bar{p}$, (c) $K^{ch}$, and (d) $\Lambda +
\Sigma $.
Right: Similar distributions, but for Pb + Pb collisions at
$\sqrt{s}=5.5$\,ATeV; (e),(f),(g),(h) correspond to (a),(b),(c)
and (d), respectively.
\label{fig2}
}
\end{figure}

What are the consequences of such peculiarity in the behavior of 
elliptic flow for the heavy-ion collisions at LHC energies? Will the
hadronic elliptic flow extend its strength or not? To answer these 
questions we generated 1,000,000 Au + Au collisions at $\sqrt{s} = 
200$\,AGeV and 500,000 Pb + Pb collisions at $\sqrt{s}=5.5$\,ATeV,
both reactions at fixed centrality $c = 42\%$. The centrality was
chosen to observe already a strong elliptic flow and, simultaneously, 
to keep the flow fluctuations at a rather modest level.
Transverse momentum dependence of elliptic flow of charged pions, 
(anti)protons, charged kaons, and $\Lambda + \Sigma$ is displayed 
in Fig.~\ref{fig2} for Au + Au and Pb + Pb collisions at RHIC (left
panels) and LHC (right panels) energies, respectively.
One can see that the flow of the most abundant hadron species at RHIC 
is stronger than its LHC counterpart. Although the soft hydro parts 
of the elliptic flow at RHIC and at LHC are similar 
\cite{hyd_v2_prc09}, the hard parts are not. More particles are 
produced at LHC via the jet fragmentation. Therefore, hadrons from
jets start to dominate over the hadrons, originated from the soft 
processes, at lower $p_T$ as compared with RHIC. The resulting 
elliptic flow at LHC is thus severely reduced. Another fact 
distinctly seen in Fig.~\ref{fig2} is the influence of the resonance 
decays on the elliptic flow of directly produced particles. Note that
according to the model calculations, resonances contribute to the
production of 80\% of pions, 70\% of protons, 60\% of $\Lambda + 
\Sigma$, and more than 50\% of kaons at LHC.
For all hadrons, except kaons at LHC, final state interactions (FSI)
increase the particle flow in the interval $1.5\,{\rm GeV}/c \leq
p_T \leq 6\,{\rm GeV}/c$. However, decays of resonances diminish the 
pion flow at $p_T \geq 1.5$\,GeV/$c$ for both energies in question.

\begin{figure}[htb]
\begin{minipage}[t]{67mm}
\epsfig{file=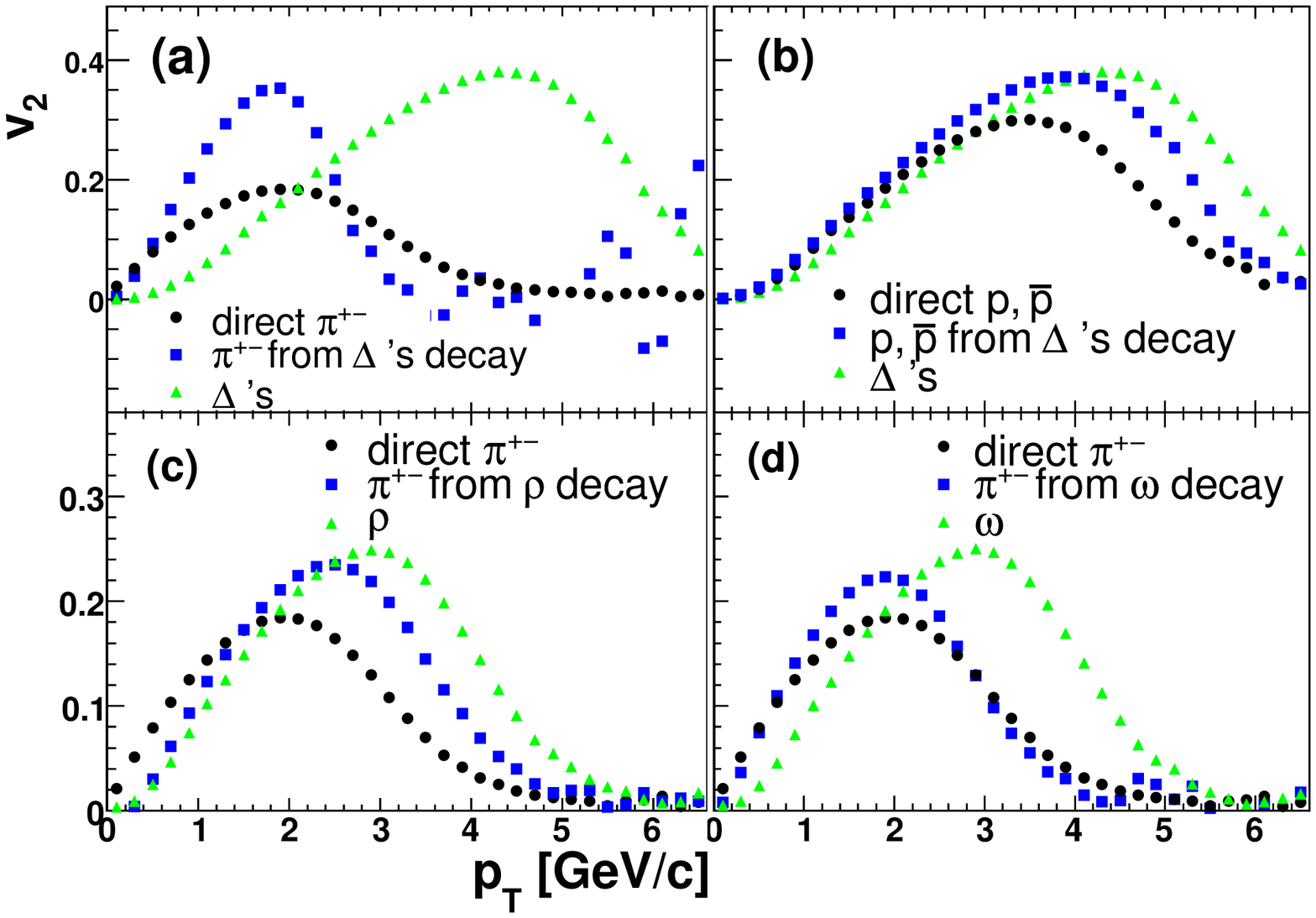,width=67mm}
\end{minipage}
\hspace{\fill}
\begin{minipage}[t]{67mm}
\epsfig{file=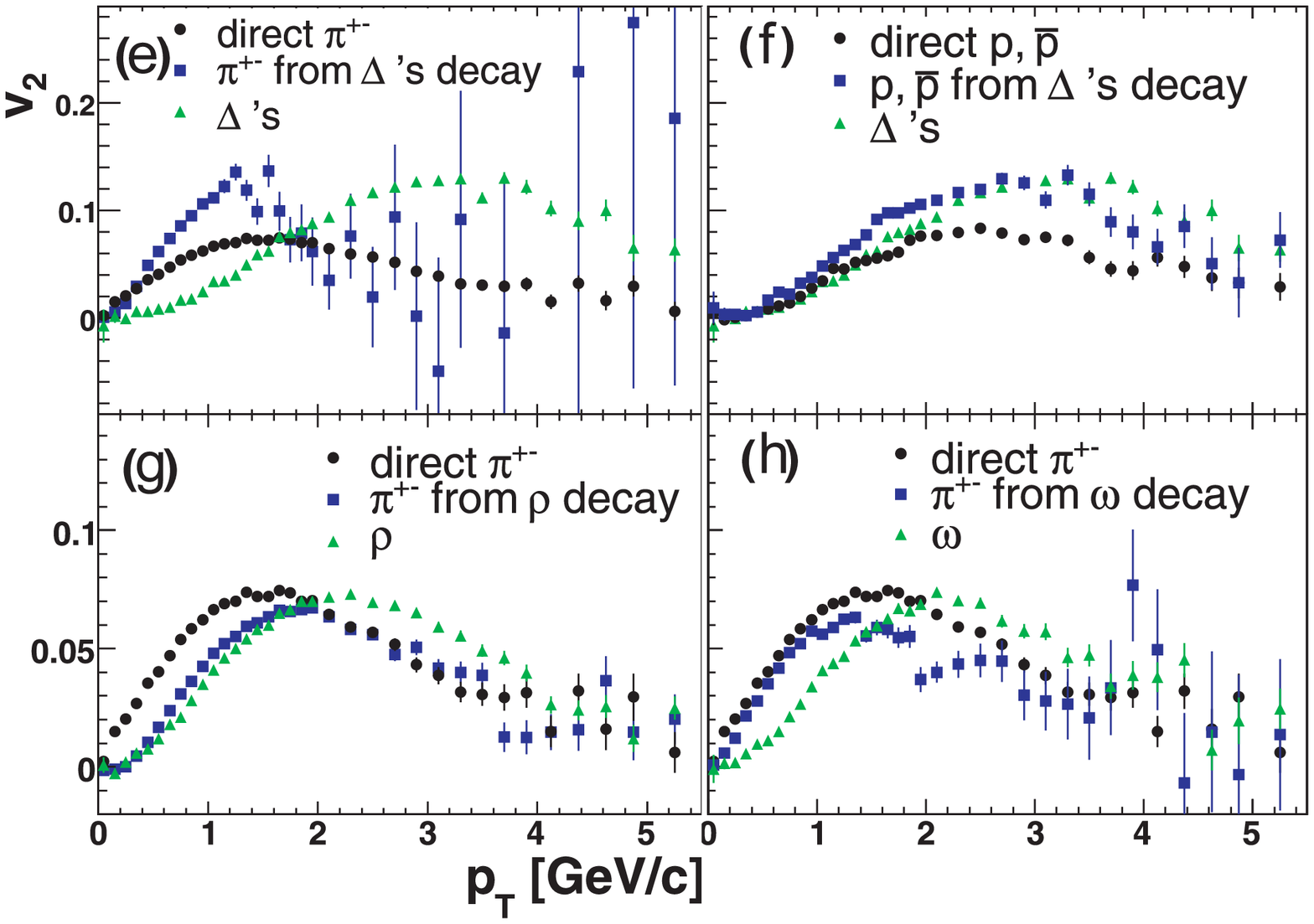,width=67mm}
\end{minipage}
\caption{Left: 
The $v_2(p_T)$ dependence for (a) $\pi^{ch}$ and (b) $p + \bar{p}$ 
produced both directly (circles) and in $\Delta$ decays (squares) in the 
HYDJET++ simulations of Au + Au at $\sqrt{s}=200$\,AGeV with 
centrality $c = 42\%$. (c) and (d) The same as (a) but for charged pions 
produced in decays of $\rho$ and $\omega$, respectively. The flow of 
resonances is shown by triangles.
Right: Similar distributions, but for Pb + Pb collisions at 
$\sqrt{s}=5.5$\,ATeV; (e),(f),(g),(h) correspond to (a),(b),(c)
and (d), respectively.
}
\label{fig3}
\end{figure}

The contributions of $\Delta$ decays to the flows of pions and 
protons, and $\rho$ and $\omega$ decays to pion elliptic flow are
presented in Fig.~\ref{fig3}. Elliptic flows of all resonances are
weaker than that of pions at low transverse momenta and stronger
at $p_T \geq 1.5$\,GeV/$c$. Due to the decay kinematics, pions from 
the decay $\omega \to 3 \pi$ are getting softer $p_T$ distributions
compared with that of $\omega$. In contrast, the transverse momentum
distributions of pions from the $\rho \to 2 \pi$ decay are close to 
that of the $\rho$ meson. As a result, elliptic flow of pions from 
the $\omega$ decay is transferred to the softer $p_T$ region compared
with $v_2^{\omega}(p_T)$, whereas elliptic flow of pions from the 
$\rho$ decay is similar to $v_2^{\rho}(p_T)$. Decays of $\Delta$'s 
produce protons with transverse momenta close to that of the mother 
particles and pions with rather soft $p_T$ spectra. Again, the 
elliptic flow of pions is shifted to softer $p_T$ region. Therefore,
the resulting elliptic flow of pions is a bit lower than the $v_2$ of
directly produced pions at $p_T \geq 1.5$\,GeV/$c$ and stronger at
intermediate transverse momenta. Modification of the elliptic flow of 
pions and light baryons can lead to violation of the hydro-induced 
mass hierarchy in the $v_2(p_T)$ sector. Particularly, the so-called 
number-of-constituent-quark (NCQ) scaling \cite{ncq_phenix,ncq_star}
can be violated. Below we check this hypothesis.   

\section{Violation of NCQ scaling}
\label{sec3}
 
\begin{figure}[htb]
\begin{minipage}[t]{67mm}
\epsfig{file=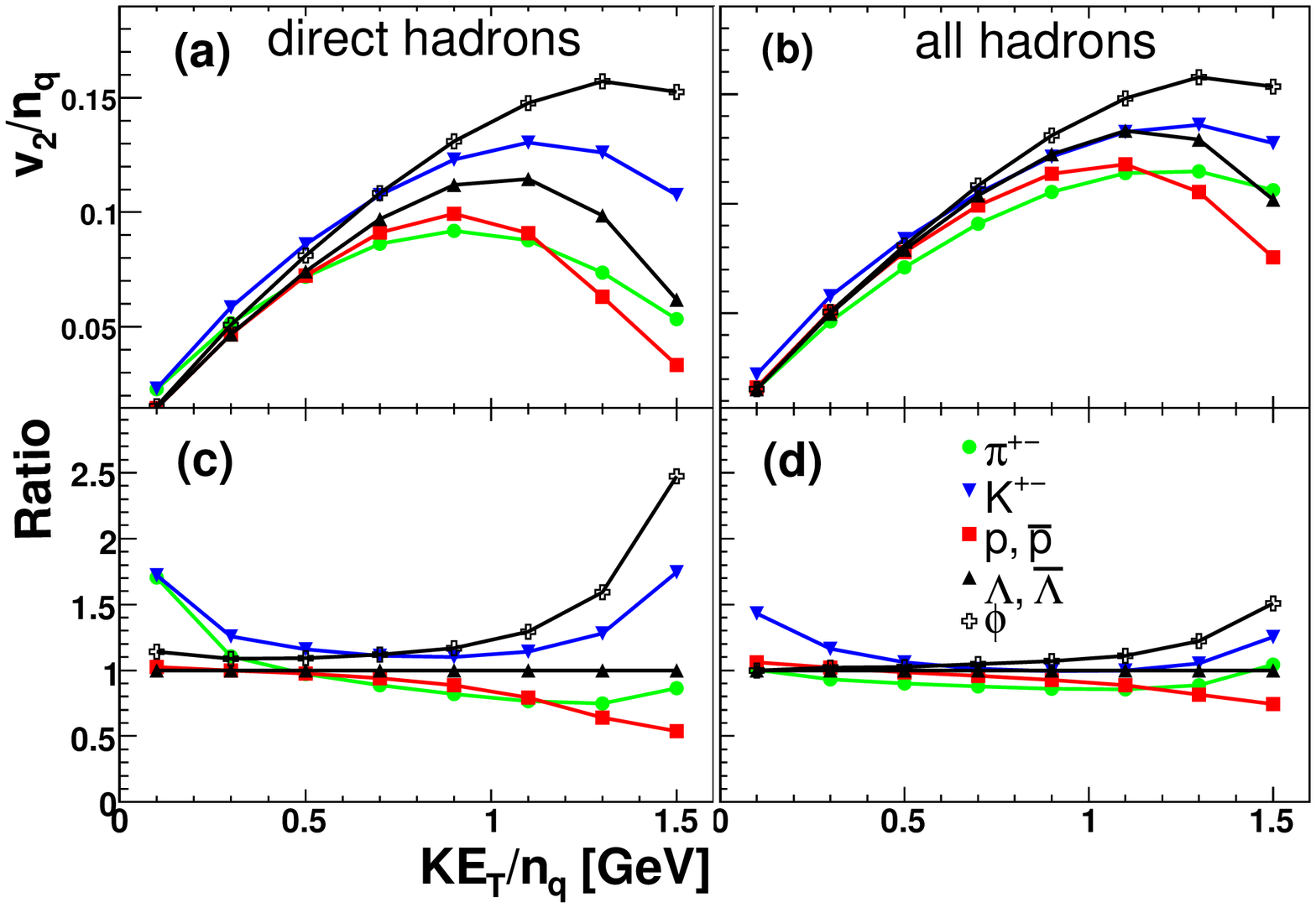,width=67mm}
\end{minipage}
\hspace{\fill}
\begin{minipage}[t]{67mm}
\epsfig{file=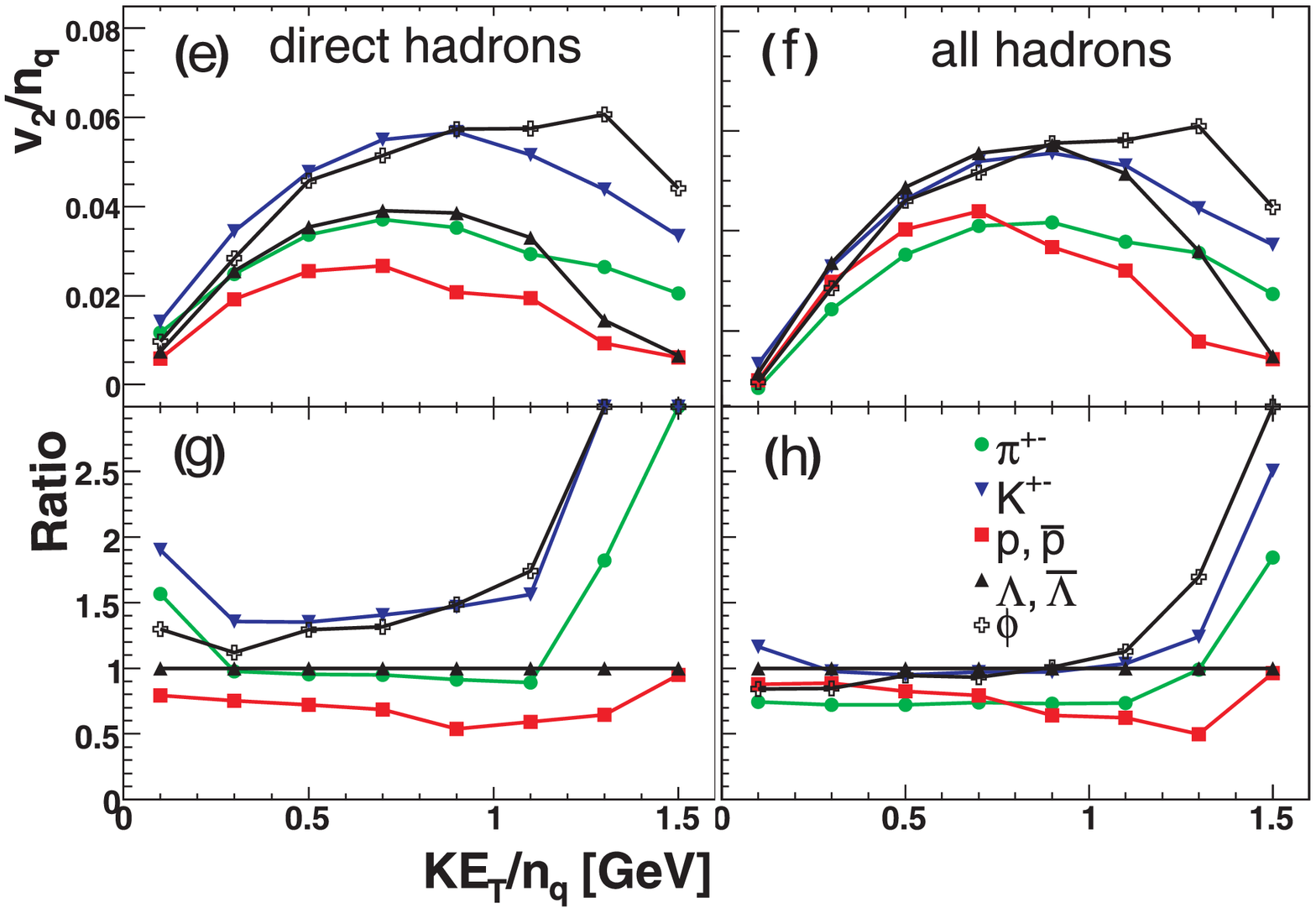,width=67mm}
\end{minipage}
\caption{Left:
Upper row: The $KE_T/n_q$ dependence of
elliptic flow for (a) direct hadrons and (b) all hadrons produced in
the HYDJET++ simulations of Au + Au at $\sqrt{s} =
200$\,AGeV with centrality $c = 42\%$. Bottom row: The $KE_T/n_q$
dependence of the ratios $(v_2/n_q)\left/(v_2^\Lambda/3) \right.$ for
(c) direct hadrons and (d) all hadrons.
Right: Similar distributions but for Pb + Pb collisions 
at $\sqrt{s}=5.5$\,ATeV; (e),(f),(g),(h) correspond to (a),(b),(c)
and (d), respectively.
}
\label{fig4}
\end{figure}

The NCQ scaling claims the similarity of the $v_2/n_q(KE_T/n_q)$
distributions, where both the elliptic flow $v_2$ and the transverse
kinetic energy of a hadron $KE_T$ are divided by the number of 
constituent quarks $n_q$. This scaling holds up until $K E_T / n_q
\approx 1$\,GeV \cite{PHENIX}. The HYDJET++ calculations of the
$v_2/n_q(KE_T/n_q)$ in heavy-ion collisions at RHIC and LHC are shown 
in Fig.~\ref{fig4} both for directly produced particles and for all
hadrons. One can see that for direct hadrons the NCQ scaling is
approximately satisfied at $K E_T / n_q \leq 1$\,GeV at RHIC only.
At LHC the NCQ scaling is not achieved for the direct particles due
to the strong influence of the jets. To see possible deviations more
distinctly, the particle distributions are also normalized in 
Fig.~\ref{fig4} to the elliptic flow of lambdas, $v_2^h/n_q^h : 
v_2^\Lambda/3$.

The decays of resonances essentially improve the situation. For 
instance, the elliptic flow of $\phi$ mesons has no resonance 
feed-down, whereas the flows of protons and lambdas are significantly
enhanced by heavy resonances at $K E_T \geq 0.5$\,GeV. Because of the
resonance decays, the NCQ scaling holds at 10\% up to $K E_T \geq
1.0$\,GeV for RHIC energies. At LHC the effect of jets is too strong,
and only the approximate number-of-constituent-quark scaling is
realized.

\section{Conclusions}
\label{sec4}

In conclusion, the interplay between hydrodynamic processes and jets,
and its influence on the formation of the particle elliptic flow is
studied within the HYDJET++ model for heavy-ion collisions at RHIC and 
LHC energies. Jet particles that carry quite weak elliptic flow start 
to dominate over the hydrolike particles at a certain transverse
momentum, thus causing the reduction of the combined $v_2$. Therefore, 
we predict that elliptic flow at LHC will be smaller than that at
RHIC, $v_2^{\rm LHC}(p_T) < v_2^{\rm RHIC}(p_T)$, at 
$p_T \geq 3$\,GeV/$c$.

Besides, jets also account for reversing of the mass ordering of the
elliptic flow at intermediate and high transverse momenta: Here the 
heaviest hadrons possess the largest flow.
Finally, the decays of resonances are shown to push the hadron 
excitation functions $v_2 / n_q (K E_T/n_q)$ toward the fulfillment 
of the NCQ scaling. However, elliptic flow of jet particles at the LHC 
energy should interfere with its hydrolike counterpart already at
intermediate transverse momenta. Thus, the realization of the
approximate number-of-constituent-quark scaling becomes worse compared 
with the RHIC case. 

{\it Acknowledgments\/.}
This work was supported in part by the QUOTA Program, Norwegian
Research Council (NFR) under contract No. 185664/V30, the
Russian Foundation for Basic Research (Grant Nos. 08-02-91001 and
08-02-92496), Grant Nos. 107.2008.2 and 1456.2008.2 of the President 
of the Russian Federation, and the Dynasty Foundation.


\end{document}